\documentclass[prb,aps,twocolumn,showpacs,amsmath,amssymb]{revtex4-1}
\usepackage{graphicx}
\usepackage{dcolumn}
\usepackage{color}

\begin{document}

\title {Phonon bottleneck in graphene-based Josephson junctions at millikelvin temperatures}

\author {I.V. Borzenets,$^{1}$ U.C. Coskun,$^{1}$ H.T. Mebrahtu,$^{1}$ Yu.V. Bomze,$^{1}$ A.I. Smirnov,$^{2}$ and G. Finkelstein$^{1}$}

\affiliation{
$^{1}$Department of Physics, Duke University, Durham, NC 27708 and \\
$^{2}$Department of Chemistry, North Carolina State University, Raleigh, NC 27695
 }

\begin{abstract}
We examine the nature of the transitions between the normal and the superconducting branches of superconductor-graphene-superconductor Josephson junctions. We attribute the hysteresis between the switching (superconducting to normal) and retrapping (normal to superconducting) transitions to electron overheating. In particular, we demonstrate that the retrapping current corresponds to the critical current at an elevated temperature, where the heating is caused by the retrapping current itself. The superconducting gap in the leads suppresses the hot electron outflow, allowing us to further study electron thermalization by phonons at low temperatures ($T \lesssim 1$K). The relationship between the applied power and the electron temperature was found to be $P\propto T^3$, which we argue is consistent with cooling due to electron-phonon interactions.

\end{abstract}


\maketitle

The electron-phonon interaction and the thermal properties of graphene have attracted a lot of attention (for an extensive review, see \emph{e.g.} Ref. \onlinecite{Balandin}.) Most studies focused on relatively high temperatures; however, this year several experiments were performed at temperatures below 10 Kelvin~\cite{Voutilainen,Vora,Yan,Fong,Betz,Price}. Measuring the intrinsic thermal properties of graphene becomes more challenging in this regime. In particular, making normal metal contacts to graphene would provide the dominant thermalization path, effectively shunting the electron cooling by phonons \cite{Price}. In order isolate the phonon contribution to thermal transport in graphene below 1K, we have contacted the graphene crystal with electrodes made from lead (Pb), which becomes superconducting below $T \sim 7$K. Superconducting leads exponentially suppress the thermal transport of hot electrons outside of the sample, allowing one to study thermalization by phonons.  

We study several superconductor-graphene-superconductor (SGS) junctions, and focus on the hysteresis between the switching current $I_S$ (from the superconducting to the normal state) and the retrapping current $I_R$ (from the normal to the superconducting state), which is observed despite the fact that our junctions are overdamped \cite{tinkham_book}. We demonstrate that the hysteresis is due to electron overheating \cite{Pekola_08,Albert_11}. Multiple contacts are made to the same graphene crystal, and resistive heating is applied at one location; simultaneously, we measure the electron temperature by monitoring the change in the critical current through a pair of contacts at a different location on the crystal. By changing the pairs of contacts where heating is applied and where the critical current is measured, we conclude that the graphene crystal is well thermalized, {\emph i.e.} its electrons maintain a uniform temperature. We evaluate the electron temperature as a function of applied power, which allows us to estimate the electron-phonon scattering rate in graphene at sub-Kelvin temperatures. 

Graphene was deposited on the Si/SiO$_2$ substrate using the standard mechanical exfoliation recipe \cite{novoselov_2005} and verified to be a single layer by Raman spectroscopy\cite{Raman}. Six parallel superconducting strips (500 nm wide) were deposited on top of the graphene crystal by thermally evaporating a 4 nm contact layer of palladium (Pd) followed by 120 nm of lead (Pb)\cite{Ivan_2011}. Thereby, a total of 5 superconductor-graphene-superconductor (SGS) junctions  were created with lengths of 0.3, 0.7, 1, 1.5 and 2 $\mu$m, labeled 1 through 5, respectively (Fig. 1a, b). The widths were roughly similar for all the junctions at around $\sim 5\mu$m.  
In order to measure the relatively small superconducting currents, the sample was encased in a copper box thermalized at the millikelvin temperature. The wires connecting the sample to the measurement setup were  filtered by cold RC filters and thermocoax cables.

At the base temperature of 35 mK the shorter junctions support a gate-dependent supercurrent (Figure 1c). The switching currents from the superconducting to normal state $I_S$ at the fully open gate voltage of $V_{gate}=40$ V were $\sim 40$ nA, $\sim 20$ nA, and $\sim 5$ nA for junctions $1 - 3$ respectively. The two longest junctions (junctions $4$ and $5$) never developed a supercurrent. These values are suppressed compared to the naive estimates of the critical current $I_C \sim \Delta /R_N e$ ($R_N$ is the normal resistance of the junction) most likely due to disorder, as discussed in our earlier publication \cite{Ivan_2011}.

Nanoscale hybrid Josephson junctions tend to be underdamped due to the large capacitance shunting the junction's bonding pads to the back gate. (The damping factor $Q$ of the Josephson junction is related to the critical current $I_C$, junction capacitance $C$ and the shunting resistance $R$ as $Q=R\sqrt{2\frac{e}{\hbar}I_C C}$~\cite{tinkham_book}.) We made the leads connecting the pads to the junctions somewhat resistive (a few hundred $\Omega$) to partially isolate the junctions from the pad capacitance and to introduce dissipation.  As a result, we estimate that our junctions are in the overdamped regime \cite{Vion}. Indeed, successive current sweeps produce a very narrow distribution of the switching and retrapping currents (see inset to Figure 2b; also note that the variations of the switching current seen in Figure 1c are reproducible on successive sweeps of $V_{gate}$.) Therefore, the observed hysteresis between the switching and retrapping current cannot be due to an underdamped dynamics, and we should seek an alternative explanation. 

\begin{figure}[]
\includegraphics[width=1 \columnwidth]{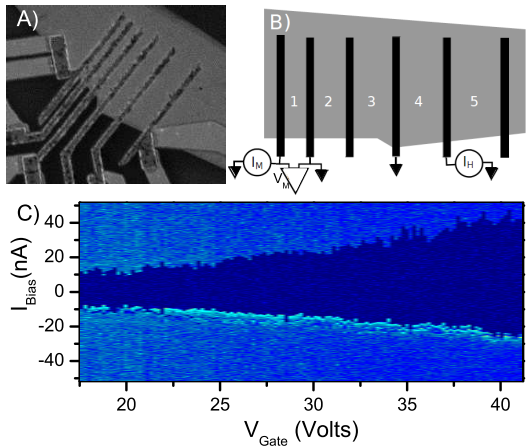}
\caption{\label{fig:overview} a) Scanning Electron Micrograph of the sample. A layer of graphene (light-colored region at the top of the image) is contacted by superconducting electrodes (dark vertical stripes) made from Pd/Pb bilayer (partially oxidized following the measurements). Six contacts form five Josephson junction of different lengths (distances between the contacts). The superconducting electrodes extend just past one of the sides of the graphene crystal, where they are contacted by the normal metal electrodes. On their other end, the superconducting electrodes do not completely cross the crystal, allowing a thermalization path to connect the different regions of graphene. b) Schematic of the measurement setup for Figure 2. Differential resistance of one junction is measured, while a heating current $I_H$ is applied to a different junction. c) Differential resistance map of junction 1 measured at the base temperature of 35 mK as a function of the back gate voltage. The supercurrent is visible as the central dark region. Notice the hysteresis between the switching (positive) and the retrapping (negative) currents. 
}
\end{figure}

In order to investigate the nature of the hysteresis, the graphene flake was heated locally by sending a current $I_H$ through one of the five junctions, while measuring the critical current of another junction~\cite{Voutilainen}. For example, junction 1 would be measured while the heating current $I_H$ is sent through junction 5 (Figure 2a). At zero heating current, the values of the switching and retrapping currents $I_S$ and $I_R$ are the largest, and both decrease as the heating power $P_H$ is increased. The difference between $I_S$ and $I_R$ also decreases and the hysteresis disappears after some intermediate value of $P_H$.

\begin{figure}[]
\includegraphics[width=1.0\columnwidth]{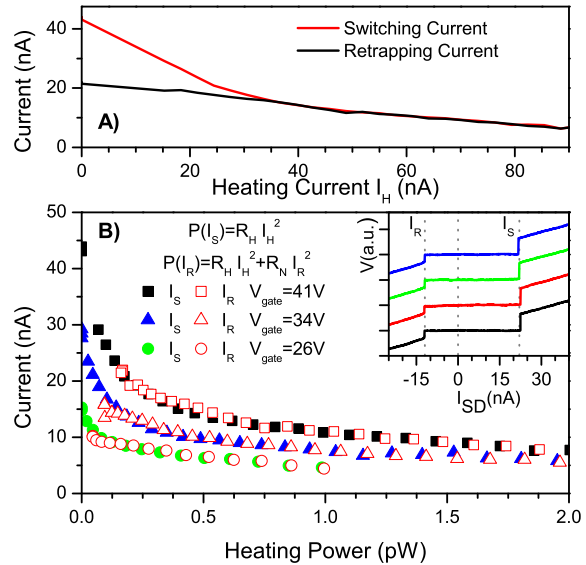}
\caption{\label{fig:hyster} 
a) The switching and the retrapping currents of the shortest junction (\#1) versus the heating current applied to the longest junction (\#5). The difference between the switching and the retrapping current disappears as more heating power is applied. b) Switching and retrapping currents versus the total heating power at different gate voltages. Since the retrapping current is necessarily measured while the junction is in the normal state, the power dissipated in that junction contributes to the total heating power. With this additional heating taken into account, the switching and the retrapping currents fall on top of each other. We conclude that the hysteresis between the switching and retrapping currents is caused by the self heating of the junction \cite{Pekola_08,Albert_11}. Inset: successive current sweeps (negative to positive) of junction 2 at $V_{gate}=40$ V. The fluctuations in values of $I_S$ and $I_R$ are negligible, indicating that the hysteresis is of a different origin compared to underdamped junctions. 
}
\end{figure}

While the switching current $I_S$ is reached when the measured junction is in the superconducting state, the retrapping current $I_R$ is realized when the measured junction is in the normal state. Therefore, at the retrapping transition, additional power $P_R=R_N I_R^2$ is applied to graphene. Taking into account this additional power, we plot $I_S (P_H)$ and $I_R(P_H+P_R)$ in Figure 2b. (Here, $P_H=R_H I_H^2$ is the power applied at the heater junction \cite{note1}.) $I_S$ and $I_R$ clearly fall on the same curve, demonstrating that at the same total dissipated power, $I_S$ and $I_R$ coincide. 

The curves of $I_S$ and $I_R$ versus $P$ map onto each other regardless of the junction that was measured (junctions 1 or 2) and for different gate voltages applied to graphene. Therefore, we can conclude that the
retrapping current is suppressed compared to the switching current due to the heat dissipated by the retrapping current itself; the hysteresis between the switching and retrapping currents is thus due to the self-heating of the measured junction. Indeed, it has been demonstrated that self-heating is the origin of the hysteretic behavior in overdamped SNS junctions \cite{Pekola_08,Albert_11}. This conclusion allows us to characterize the electron cooling in graphene in the remainder of the paper.

\begin{figure}[]
\includegraphics[width=0.8 \columnwidth]{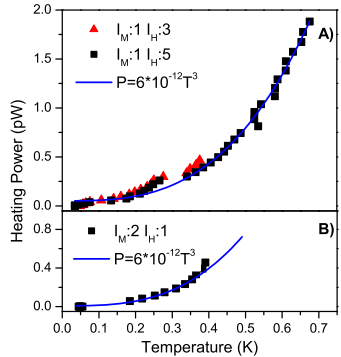}
\caption{\label{fig:power_law} Dissipated power \emph{vs.} electron temperature in graphene. The latter was extracted by matching the switching current at a given heating current to the switching current measured by heating the whole sample. The graph presents three measurement configurations: a) measuring the switching current of junction 1 while heating junction 3 or junction 5, and b) measuring junction 2 while heating junction 1. The data for all three configurations fit well to a $P\propto T^3$ power law dependence with less than $5\%$ difference in the proportionality factor, meaning that the electrons in the entire region are thermalized. 
}
\end{figure}

Instead of heating the graphene crystal locally by the current $I_H$, we can heat the entire sample. We measure the switching current $I_S$ in the two shortest junctions versus temperature $T$ at several different gate voltages. By comparing the $I_S$ \emph{vs.} $T$ data with the $I_S$ \emph{vs.} $P$ data, we are able to extract the temperature up to which the electrons in the measured junction are heated for a given $P$. This way, a plot of $P$ \emph{vs.} $T$ has been extracted for measured junction 1 while heating either junction 3 or 5 (Figure 3a). Similarly, in Fig 3b, we measured junction 2 while heating junction 1. (In the latter case, large enough current was passed through junction 1 to overcome its own supercurrent. Once that happens, graphene heats to about 200 mK, which explains the gap in the data at lower temperatures.)

The three curves in Figure 3 are remarkably close, although the three setups have different distances and configurations of the heater and the measured junctions, implying that the entire region of graphene crystal in the vicinity of the junctions is thermalized to a uniform electron temperature. Empirically, the curves can be fitted by the same power law $P=6 \cdot 10^{-12} \mathrm{\frac{W}{K^3}}~T^3$. This dependence rules out cooling by the hot electron diffusion into the superconducting leads. Indeed, the dependence would be exponentially suppressed in temperature due to the gap in the quasiparticle density of states of the superconductor. Even at $T=$1 K, the dissipated power ($6 \cdot 10^{-12}$ W) is several orders of magnitude larger than that estimated from the Wiedemann-Franz law, properly adjusted for the suppressed density of states in the leads ($\sim 10^{-16}$ W). This situation is different from the regime realized in Ref. \onlinecite{Voutilainen}, where a relatively small superconducting gap in Al leads allows them to provide a dominant thermalization path. We conclude that in our sample electrons in the graphene crystal are thermally decoupled from the leads and must be cooled by phonons. 

Furthermore, the observed $P(T)$ cannot be limited by the Kapitza resistance between graphene and the substrate, which is estimated to be a few orders of magnitude smaller. (As an upper limit for the resistance, we take the values measured at 40 K in Ref. \onlinecite{Chen_09} and extend the cooling power according to $\propto T^3$ or $\propto T^4$.) Hence, graphene lattice must be well thermalized with the substrate. We conclude that the bottleneck for cooling in our system is the weak thermal coupling between electrons and phonons in graphene.

Following the derivation of Ref.~\onlinecite{DasSarma_08}, we estimate the electron-phonon cooling power in graphene as $\frac{A D^2 \sqrt n}{\hbar^3 \rho_m s^2 v_F^2 l} (k_B T_{el})^3 $. Here, $A$ is the area of graphene, $D$ is the deformation potential constant, $n$ is the electron density, $\rho_m$ is graphene mass density, $s$ is the speed of sound, and $T_{el}$ is the electron temperature. We assume that at low temperatures the wavelength of the emitted phonon $hs / k_B T_{el}$, which enters the scattering matrix element, should be replaced by a distance $l$ of the order of a hundred nm (\emph{e.g.} the distance between the leads, or the electron mean free path). We also took into account that Ref. \onlinecite{DasSarma_08} calculated the transport time while we are interested in the scattering time, because each phonon emission event typically results in the electron energy loss of the order of $k_B T_{el}$.

Plugging in the numbers from Refs. \onlinecite{DasSarma_08,Kim_10}, we estimate the cooling power at $P=10^{-12} \mathrm{\frac{W}{K^3}}~T_{el}^3$, within an order of magnitude from the measured value. [This is a rather crude estimate, in part because we do not know how large is the thermalized area of graphene crystal, which extends beyond the immediate vicinity of the contacts, see Figure 1a. This area determines $A$, taken to be $(10 \mu$m$)^2$ in this estimate.] 

Recent measurements of Refs. \onlinecite{Fong,Betz} report cooling power for hot electron $P \propto T_{el}^4$, indicating that the emitted phonon wavelength $hs / k_B T_{el}$ is not cut off, as we assumed above. The difference may stem both from the lower temperatures in our measurement and a more restricted sample geometry resulting in shorter $l$ in our case. The cooling power of Ref. \onlinecite{Fong}, extrapolated down to 1 K would yield 0.07 W/m$^2$, which in fact is close to our result for $A=(10 \mu$m$)^2$. 

Finally, $P \propto T_{el}^3$ dependence due to supercollision cooling \cite{Levitov} has been just observed in Refs. \onlinecite{Betz2,Graham}. This regime is realized if the wavevector of the emitted phonon, $k_B T_{el}/s$, exceeds the electron Fermi wavevector $k_F$; the emission process is enabled by the disorder.  We work in the opposite regime, $k_F \gg k_B T_{el}/s$, so that the theory of Ref. \onlinecite{Levitov} cannot be directly applied.

In conclusion, we have shown that the difference between the switching and the retrapping currents in our graphene Josephson junctions is caused by electron overheating in the normal state. The superconducting contacts thermally isolate the graphene crystal from the leads, allowing us to measure the electron temperature rise for a given dissipation power, and hence the electron-phonon energy transfer rate. The observed power law dependence $P\propto T^3$ is consistent with theory of electron-phonon interactions in graphene and with other measurements. Due to small electron heat capacitance and their decoupling from phonons, this type of sample may be useful for detector applications \cite{Prober}.


\begin{thebibliography}{99}



\bibitem{Balandin} A. A. Balandin, Nature Materials, {\bf 19}, 569 (2011). 

\bibitem{Voutilainen} J. Voutilainen, A. Fay, P. H\"{a}kkinen, J. K. Viljas, T. T. Heikkil\"{a}, and P. J. Hakonen, Phys. Rev. B {\bf 84}, 045419 (2011).

\bibitem{Betz} A. C. Betz, F. Vialla, D. Brunel, C. Voisin, M. Picher, A. Cavanna, A. Madouri, G. F`eve, J.-M. Berroir, B. Pla\c{c}ais, and E. Pallecchi, Phys. Rev. Lett. {\bf 109}, 056805 (2012).

\bibitem{Fong} K. Fong and K. Schwab, Phys. Rev. X {\bf2}, 4 (2012).

\bibitem{Vora} H. Vora, P. Kumaravadivel, B. Nielsen, and X. Du, Appl. Phys. Lett. {\bf100}, 153507 (2012).

\bibitem{Yan} J. Yan, M.-H. Kim, J. A. Elle, A. B. Sushkov, G. S. Jenkins, H. M. Milchberg, M. S. Fuhrer, and H. D. Drew, Nature Nanotech. {\bf7}, 472 (2012).

\bibitem{Price} A. S. Price, S. M. Hornett, A. V. Shytov, E. Hendry, and D. W. Horsell, Phys. Rev. B {\bf 85}, 161411(2012). 

\bibitem{tinkham_book} M. Tinkham, \emph{Introduction To Superconductivity} (McGraw-Hill, 1996).


\bibitem{Pekola_08} H. Courtois, M. Meschke, J. T. Peltonen,  and J. P. Pekola, Phys. Rev. Lett. {\bf 101}, 067002 (2008).

\bibitem{Albert_11} P. Li, P. M. Wu, Y. V. Bomze, I. V. Borzenets, G. Finkelstein, A. M. Chang, Phys. Rev. Lett. {\bf 107} 137004 (2011) and Phys. Rev. B {\bf 84}, 184508 (2011). 

\bibitem{novoselov_2005} K. S. Novoselov, D.  Jiang, F. Schedin, T.J. Booth, V. V. Khotkevich, S. V. Morozov, and A. K. Geim, PNAS {\bf 102}, 10451 (2005). 

\bibitem{Raman} A. C. Ferrari, J. C. Meyer, V. Scardaci, C. Casiraghi, M. Lazzeri, F. Mauri, S. Piscanec, D. Jiang, K. S. Novoselov, S. Roth, and A. K. Geim, Phys. Rev. Lett. {\bf 97} 187401, (2006).

\bibitem{Ivan_2011} I. V. Borzenets, U. C. Coskun, S. J. Jones, and G. Finkelstein, Phys. Rev. Lett. {\bf 107}, 137005 (2011) and IEEE Trans. Appl. Supercond. {\bf 22}, 1800104 (2012).

\bibitem{Vion} D. Vion, M. G\"{o}tz, P. Joyez, D. Esteve, and M. H. Devoret, Phys. Rev. Lett.,  {\bf77},16, 3435, (1996).

\bibitem{note1} We used $R_H$ as a fitting parameter to ensure the best match between the $I_S (P_H)$ and $I_R(P_H+P_R)$ curves. The resulting values of $R_H$ are close to the measured two-probe resistances of the heater junctions. The difference between the two-probe resistance and $R_H$ is explained by a more sophisticated set-up used in the heating measurement, where a fraction of the heating current can flow to several of the grounded contacts other than the designated heater ground. These parallel current paths reduce the effective resistance of the heater. 

\bibitem{Chen_09} Z. Chen, W. Jang, W. Bao, C. N. Lau, and C. Dames, Appl. Phys. Lett. {\bf 95}, 161910 (2009). 

\bibitem{DasSarma_08} E. H. Hwang and S. Das Sarma, Phys. Rev. B {\bf 77}, 115449 (2008). 

\bibitem{Kim_10} D. K. Efetov and P. Kim, Phys. Rev. Lett {\bf 105}, 256805 (2010).  

\bibitem{Levitov} J. C. W. Song, M. Y. Reizer, and L. S. Levitov, Phys. Rev. Lett. {\bf 109}, 106602 (2012). 

\bibitem{Betz2} Supercollision cooling in undoped graphene

\bibitem{Graham} Photocurrent measurements of supercollision cooling in graphene

\bibitem{Prober} C.B. McKitterick, D.E. Prober, and B.S. Karasik, arXiv:1210.5495v3 (2012). 

\end{thebibliography}
\end{document}